\documentclass[12pt]{revtex4-1}
\usepackage{amsmath,graphicx,hyperref,bm}
\usepackage[font={small},flushleft,indent]{caption}
\begin{document}
\title{Black hole shadow in the view of freely falling observers}

\author{Zhe Chang,}
\author{Qing-Hua Zhu}
\email{zhuqh@ihep.ac.cn}

\affiliation{Institute of High Energy Physics, Chinese Academy of Sciences, Beijing 100049, China}
\affiliation{University of Chinese Academy of Sciences, Beijing 100049, China}



\begin{abstract}
	First sketch of black hole from M87 galaxy was obtained by Event Horizon Telescope, recently. As appearance of black hole shadow reflects space-time geometry of black holes, observations of black hole shadow may be a promising way to test general relativity in strong field regime. In this paper, we focus on angular radius of spherical black hole shadow with respect to freely falling observers. In the framework of general relativity, aberration formulation and angular radius-gravitational redshift relation are presented. For the sake of intuitive, we consider parametrized Schwarzschild black hole and Schwarzschild-de Sitter black hole as representative examples. We find that the freely in-falling observers would observe finite size of shadow, when they go through inner horizon. For observers freely falling from the outer horizon of Schwarzschild-de Sitter black hole, we find that the angular radius of the shadows could increase even when the observers move farther from the black hole.
\end{abstract}

\maketitle

{\section{Introduction}

Event Horizon Telescope has recently obtained first sketch of black hole from M87 galaxy \cite{collaboration_first_2019}. Investigation of black hole might come up to an observation era in the near future. Besides gravitational wave from coalescence of binary black hole or compact stars \cite{Abbott:2016nmj,TheLIGOScientific:2017qsa},
the directed observation of black hole can be another way to test general relativity in strong gravity regime.

In strong gravitational field of black hole, even light can not escape out of black hole from event horizon. In order to find an observable of black hole, Synge \cite{synge_escape_1966} focused on bending rays around black hole. He firstly considered a Schwarzschild black hole that would catch rays from outside of it. The bending rays can escape to spatial infinity again, only if the rays do not reach $\frac{3}{2}$ times of Schwarzschild radius $r_s$ from centre of black hole. The critical surface $r= \frac{3}{2} r_s$ of Schwarzschild space-time is called photon sphere. We present a schematic diagram of the black hole shadow in Figure~\ref{Fig-1}. For those rays from outside light sources without approaching the photon sphere, we can observe them in the field of view $\rm II$. While, in the field of view $\rm I$,  we would observe nothing but darkness. All these dark region in the view is what we call shadow of black hole. 
\begin{figure}[ht]
	\centering
	{\includegraphics[width=0.7\linewidth]{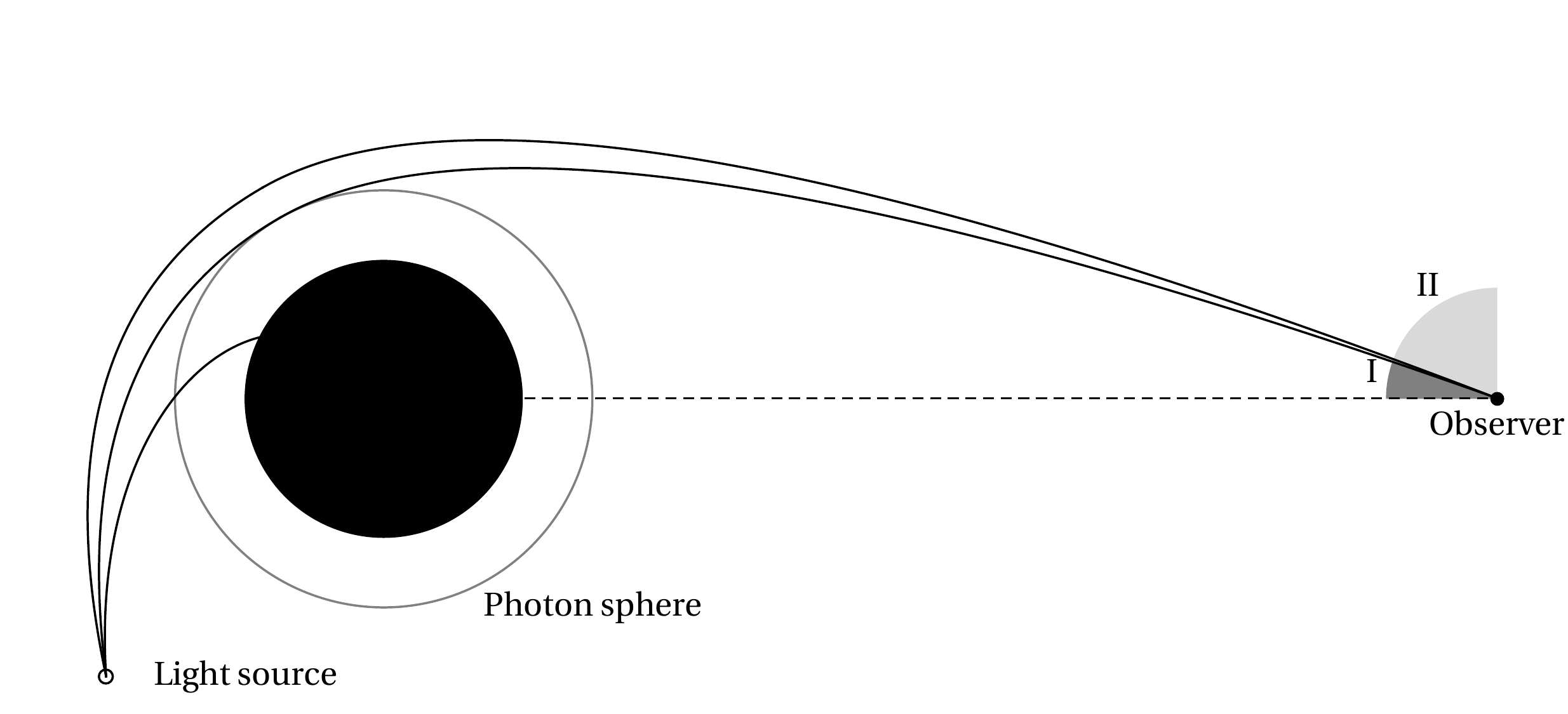}}
	\caption{Schematic diagram of black hole shadow studied by Synge \cite{synge_escape_1966}. In field of view $\rm I$, we would observe nothing but darkness. All these dark region in the view constitute what we called black hole shadow. \label{Fig-1}}
\end{figure}

In fact, size and shape of shadow can reflect space-time geometry of black holes. Since the pioneer work of Synge \cite{synge_escape_1966}, investigation of black hole shadow has been extended in various kinds of black holes. Shadow of Kerr black hole was firstly studied by Bardeen \cite{bardeen_timelike_1973}. And recent works have gone further to consider extended Kerr black holes, such as deformed black hole \cite{atamurotov_shadow_2013}, Kerr-Newman-NUT--(anti--)de Sitter black hole \cite{grenzebach_photon_2014}, regular black hole  \cite{li_measuring_2014,abdujabbarov_shadow_2016}, accelerated Kerr black hole \cite{grenzebach_photon_2015} and Kerr black hole with scalar hair \cite{cunha_shadows_2015} or dilation parameter \cite{wei_observing_2013}. From model independent methods, there were also developments of general parametrization of black holes  \cite{johannsen_metric_2011,johannsen_photon_2013,rezzolla_new_2014,konoplya_general_2016,younsi_new_2016,tian_testing_2019}. These parametrized black holes could also provide parameters space for modified theory of gravity \cite{ayzenberg_black_2018}. Besides extending parameter space of  black hole, physically or phenomenally, the observation of shadow also could be used to constrain dark matter \cite{davoudiasl_ultralight_2019,haroon_shadow_2019} or dark halo \cite{konoplya_shadow_2019} models. As it's suggested that shadow of a compact object was not necessarily a black hole \cite{cunha_does_2018}, more aggressive studies have already referred to shadow of naked singularity \cite{hioki_measurement_2009,ortiz_shadow_2015} and wormhole \cite{ohgami_wormhole_2015,ohgami_wormhole_2016}.

On the other side, the appearance of black hole shadow is also related to how we observe it, such as inclination angle of observation \cite{hioki_measurement_2009,tsupko_analytical_2017} and motional status of observers \cite{grenzebach_aberrational_2015}. 
Especially, as the universe is expanding, it's well-motivated to study how the expanding rate affects observed size or shape of black hole shadow. Recently, Perlick, Tsupko and Bisnovatyi-Kogan \cite{perlick_black_2018,bisnovatyi-kogan_shadow_2018} considered this via finding a co-moving observer driven by the expansion of the universe. In large distance, it indicates how expanding rate affects size of black hole shadow. While, in near region, the co-moving observers seem to be less meaningful, as it's hard for an observer to be unaffected by gravity of the black hole.

As for reference frame with respect to a motional observer, it seems beyond description of general relativity to date. Local reference frame \cite{misner_gravitation_1973, mashhoon_hypothesis_1990,gourgoulhon_auth._special_2013}  seems partly solving the problem of arbitrary motional reference frames. It could be used to resolve frame dragging and Lense--Thirring effect \cite{misner_gravitation_1973} and calculate the shadows of rotating black holes \cite{bardeen_timelike_1973,grenzebach_photon_2014}, while, it not suitable for dealing with the Unruh effect \cite{davies_scalar_1975,unruh_notes_1976}. 
From alternative perspectives of the general covariance, we can consider a well-defined motional reference frame that should satisfy some trivial properties as follows. Firstly, different motional status of reference frame should be related with coordinate transformation of space-time, $X^{\mu} = X^{\mu} (x^{\nu})$. Secondly, the coordinate time of a motional reference frame should adapt to 4-velocity of itself, namely adapted coordinate, $- N {\rm{d}} T = u_{\mu} {\rm{d}} x^{\mu}$. It comes from the first step of 3+1 foliation. Thirdly, the motional reference frame should regard itself as static frame, namely $\frac{{\rm{d}} X^i}{{\rm{d}} \tau} = 	u^{\mu} \partial_{\mu} X^i = 0$. One can find that there is a robust instance that satisfies these properties in Schwarzschild space-time, namely, Lemaitre coordinate \cite{lemaitre_expanding_1997}.

In this paper, we consider black hole shadow with respect to freely falling observers. In Schwarzschild space-time, Lemaitre coordinate describes the reference frame of freely falling observers from spatial infinity. It can be generalized in the space-time of Schwarzschild-like spherical black hole for observers freely falling from finite distance \cite{misner_gravitation_1973}. In this coordinate, we derive aberration formulation. It can show how size of shadow is affected by motion status of reference frame. For the sake of intuitive, we consider parametrized Schwarzschild black hole proposed by Tian \cite{tian_testing_2019} and Schwarzschild-de Sitter black hole as representative examples. The in-falling observers would observe finite size of shadow, when they go through the inner horizon. It's different from that with respect to static observers \cite{synge_escape_1966}. For Schwarzschild-de Sitter black hole, the space-time is not asymptotic. One can use the cosmological constant of the black hole to estimate how much effect the expanding rate of the universe would take on the size of shadow. We find the angular radius of shadow could increase at the beginning when observers move farther from the black holes.

This paper is organized as follows, In section \ref{II}, we review aberration formulation in special relativity. In section \ref{III}, we introduce generalized Lemaitre coordinate and calculate corresponding angular radius of spherical black hole shadow. It shows that aberration formulation could be the form of that in special relativity, if the velocities of observers and light rays are expressed in term of orthogonal tetrads. 
For the sake of intuitive, in the final parts of section \ref{III}, \ref{IV}, we use parametrized Schwarzschild black hole and Schwarzschild-de Sitter black hole as examples. In section \ref{V}, we study time evolution of angular radius of shadow in the view of freely falling observers. Finally, conclusions and discussions are summarized in section \ref{VI}.

\section{Aberration formulation in flat space-time}\label{II}

As speed of light is finite, apparent angular position of an object would depend on velocity of observers \cite{penrose_apparent_1959} and should be understood as aberration effect. In this section, we would review derivation of aberration formulation in flat space-time.

The metric of flat space-time is given by
\begin{equation}
	{\rm{d}} s^2 = - {\rm{d}} t^2 + {\rm{d}} r^2 + r^2 {\rm{d}} \Omega^2~.
\end{equation}
We can calculate null geodesic equations from
\begin{eqnarray}
	E & = & \frac{{\rm{d}} t}{{\rm{d}} \lambda}~,\\
	L & = & r^2 \frac{{\rm{d}} \phi}{{\rm{d}} \lambda}~,\\
	0 & = & - \left( \frac{{\rm{d}} t}{{\rm{d}} \lambda} \right)^2 + \left(
	\frac{{\rm{d}} r}{{\rm{d}} \lambda} \right)^2 + r^2 \left( \frac{{\rm{d}}
		\phi}{{\rm{d}} \lambda} \right)^2~,
\end{eqnarray}
where $E$ and $L$ are integral constants of geodesic equations. From the 4-velocity of null curves, one can obtain
\begin{equation}
	\frac{{\rm{d}} r}{{\rm{d}} \phi} = r^2 \sqrt{\left( \frac{E}{L} \right)^2 -
		\frac{1}{r^2}}~.
\end{equation}
Then, angular position $\psi_{ \rm{stat}}$ is given by
\begin{equation}
	\cot \psi_{ \rm{stat}} = \frac{{\rm{d}} r}{r {\rm{d}} \phi} = r \sqrt{\left(
		\frac{E}{L} \right)^2 - \frac{1}{r^2}} \label{6}~.
\end{equation}

On the other side, we turn to consider the angular position with respect to a uniformly motional reference frame $(T, R, \Theta, \Phi)$. For simplicity, we consider the reference frame $(T, R, \Theta, \Phi)$ that undergoes a constant 3-velocity $\upsilon$ along radial direction with respect to the reference frame $(t, r, \theta, \phi)$. These reference frames are related by the Lorentz transformation,
\begin{equation}
	\left\{\begin{array}{lll}
		T      & = & \frac{t}{\sqrt{1 - \upsilon^2}} - \frac{\upsilon r}{\sqrt{1 -\upsilon^2}}~, \\
		R      & = & - \frac{\upsilon   t}{\sqrt{1 - \upsilon^2}} +
		\frac{r}{\sqrt{1 - \upsilon^2}}~,                                                        \\
		\Theta & = & \theta~,                                                                    \\
		\Phi   & = & \phi~.
	\end{array}\right.
\end{equation}
From the transformation and null geodesic, one can obtain
\begin{equation}
	\frac{{\rm{d}} R}{{\rm{d}} \Phi} = \frac{r^2}{\sqrt{1 - \upsilon^2}} \left( -
	\upsilon \left( \frac{E}{L} \right) + \sqrt{\left( \frac{E}{L} \right)^2 -
		\frac{1}{r^2}} \right)~.
\end{equation}
Therefore, angular position $\psi_{ \rm{mov}}$ observed by a uniformly motional observer is
given by
\begin{equation}
	\cot \psi_{ \rm{mov}} = \frac{{\rm{d}} R}{r (T, R) {\rm{d}} \Phi} = \frac{r}{\sqrt{1 - \upsilon^2}} \left( - \upsilon \left( \frac{E}{L} 	\right) + \sqrt{\left( \frac{E}{L} \right)^2 - \frac{1}{r^2}} \right)~. \label{9}
\end{equation}

From Eqs.~(\ref{6}) and (\ref{9}), the relation between $\psi_{ \rm{stat}}$ and
$\psi_{ \rm{mov}}$ is read as aberration formulation,
\begin{equation}
	\cot \psi_{ \rm{mov}} = \cot \psi_{ \rm{stat}} \left( 1 -
	\frac{\upsilon}{c_r} \right) \frac{1}{\sqrt{1 - \upsilon^2}} \label{10}~,
\end{equation}
where $c_r \equiv \frac{{\rm{d}} r}{{\rm{d}} t}$ and speed of light $c =
	\sqrt{\left( \frac{{\rm{d}} r}{{\rm{d}} t} \right)^{2} + r^2 \left( \frac{{\rm{d}}
			\phi}{{\rm{d}} t} \right)^2}$. In the case that light propagates nearly along
direction of $\upsilon$, the $c_r$ tends to be $c$. Thus, we can rewrite the
Eq.~(\ref{10}) as
\begin{equation}
	\cot \psi_{ \rm{mov}} = \cot \psi_{ \rm{stat}} \sqrt{\frac{1 -
			\upsilon}{1 + \upsilon}}~.
\end{equation}
The aberration formulation was also referred in Ref. \cite{penrose_apparent_1959,pauli_theory_2013,rindler_essential_2013}.
It can provide a physical intuition of how
angular position is affected by motional status of the reference frames.

Besides, one can find that the key of derivation of aberration formulation is how to describe coordinate transformation of different reference frames.
It can be generalized in curved space-time. We would employ the derivation on angular radius of black hole shadow in the following sections.

\section{Angular radius of spherical black hole shadow for freely-falling observers }\label{III}

In classical theory of gravity, freely falling observers would feel nothing special, when going through event horizon of black hole. It suggests that freely falling observers could observe the shadow in the interior of black hole, as light runs faster than massive test particles. However, studies referred to interior of black hole seems not to be falsifiability. In this paper, we would focus on the angular radius of shadow with respect to freely falling observers in the exterior of spherical black holes. The results are different from that with respect to a static observer in strong gravity regime.

\subsection{Generalized Lemaitre coordinate}

We start from a Schwarzschild-like spherical black hole,
\begin{eqnarray}
	{\rm{d}} s^2  =  - f (r) {\rm{d}} t^2 + \frac{1}{f (r)} {\rm{d}} r^2 + r^2
	{\rm{d}} \Omega^2~, \label{12}
\end{eqnarray}
where $f (r)$ is an arbitrary function of coordinate $r$. The form of $f (r)$ depends on theory of gravity. From Eq.~(\ref{12}), we consider radial time-like geodesic equations and simplify them as
\begin{eqnarray}
	\left\{\begin{array}{rll}
	\frac{{\rm d} t}{{\rm d}\tau} & = & \frac{\mathcal{E}}{f(r)}~, \\
	\frac{{\rm d} r}{{\rm d}\tau} & = & \pm \sqrt{\mathcal{E}^2 -f(r)}~, \\
	\frac{{\rm d} \theta}{{\rm d}\tau} &=& 0~, \\
	\frac{{\rm d} \phi}{{\rm d}\tau} &=& 0~,
	\end{array}\right. \label{13-1}
\end{eqnarray}
where we set $u^{\mu}\equiv\frac{{\rm d}x^{\mu}}{{\rm d}\tau}$. For an observer freely falling from location $r_0$, the integral constant $\mathcal{E}$ can be determined, namely, $\mathcal{E} = \sqrt{f_0} \equiv \sqrt{f(r_0)}$. 

For coordinate adapted to a free-falling observer $u^{\mu} = \left( \frac{\sqrt{f_0}}{f}, \pm \sqrt{f_0 - f}, 0, 0 \right)$, it's formulated as
\begin{equation}
	- N {\rm d} T = u_{\mu} {\rm d}x^{\mu}~, \label{14a}
\end{equation}
where $x^{\mu}$ can take $\{t,r,\theta,\phi\}$, indices of $u_\mu$ is lowered by metric Eq.~(\ref{12}), $T$ is coordinate time of the adapted coordinates and $N$ is an integrating factor to make sure ${\rm d}^2T = 0 $. In the case that $u^{\mu}$ is derived from geodesic equations, it's so-called geodesic slice and $N=1$ always exists \cite{gourgoulhon_3+1_2012}. Thus, we can rewrite Eq.~(\ref{14a}) as
\begin{equation}
	{\rm d} T = \sqrt{f_0}{\rm d}t \mp \frac{\sqrt{f_0 - f}}{f} {\rm d} r~. \label{15a}
\end{equation}
With requirement that motional reference frame should regard itself as static frame, namely, $ \frac{{\rm d} X^i}{{\rm d}\tau} =u^{\mu}\partial_{\mu}X^i = 0$, one can obtain
\begin{eqnarray}
	\left\{\begin{array}{lll}
	{\rm d}R & = & \pm \frac{{\rm d}t}{C}+\frac{{\rm d}r}{C f}\sqrt{\frac{f_0}{f_0-f}}~, \\
	{\rm d}\Theta & = & {\rm d}\theta~,\\
	{\rm d}\Phi & = & {\rm d}\phi ~,
	\end{array}\right.  \label{16a}
\end{eqnarray}
where $X^i$ can take $\{R,\Theta,\Phi\}$ and $C$ is a constant and we would explain it latter. Using Eqs.~(\ref{12}), (\ref{15a}) and (\ref{16a}), we can obtain metric in adapted coordinate $(T,R,\Theta,\Phi)$ for the freely falling observer $u^{\mu}$,
\begin{eqnarray}
	{\rm{d}} s^2  = - {\rm{d}} T^2 + C^2 (f_0 - f) {\rm{d}} R^2 + r^2 {\rm{d}}
	\Omega^2~. \label{13}
\end{eqnarray}
The explicit coordinates also can be obtained via integrals of Eqs.~(\ref{15a}) and (\ref{16a}),
\begin{equation}
	\left\{\begin{array}{lll}
		T      & = & \sqrt{f_0} (t - t_0) \mp \int^r_{r_0} {\rm{d}} r \left\{
		\frac{\sqrt{f_0 - f}}{f} \right\}~,                                          \\
		R      & = & \mp \frac{1}{C} (t - t_0) + \frac{1}{C} \int_{r_0}^r {\rm{d}} r
		\left\{ \frac{1}{f} \sqrt{\frac{f_0}{f_0 - f}} \right\}~,                    \\
		\Theta & = & \theta~,                                                        \\
		\Phi   & = & \phi~.
	\end{array}\right.  \label{14}
\end{equation}
In the adapted coordinate, the coordinates $(T,R,\Theta,\Phi)$ can describe motional reference frame for freely falling observers. Because one can find that the 4-velocity $u^{\mu}$ of the observers turns to be $u^{\mu} = (1, 0, 0, 0)$ in the reference frame.

In Schwarzschild space-time that $f (r) = 1 - \frac{2 M}{r}$, origin Lemaitre coordinate is specific case that $C = 1$ and $f_0 = 1$. One can find that the origin Lemaitre coordinate is no longer an asymptotic space-time at $r 	\rightarrow \infty$. In order to deal with this case, we introduce a normalization constant $C \equiv \left( \sqrt{f_0 - 1} \right)^{- 1}$. In the case of $r \rightarrow r_0 =   \infty$, one can obtain $T = t-t_0$ and $R = 	r - r_0$ from the transformation, Eq.~(\ref{14}). The generalized Lemaitre coordinate with normalization constant $C$ can provide an explicit meaning of coordinates $T$ and $R$.

\subsection{Angular radius of the shadows for free-falling observers}

Due to spherical symmetry of the black holes, we can set $\Theta = \theta = 	\rm{\pi/2}$ for simplicity. The calculation of  angular radius of black hole shadow is similar to the derivation of angular position in flat space. The difference is that we should consider bending light in curved space-time. Namely, trajectory of light is described by null geodesic equations. From Eq.~(\ref{13}), we can obtain 4-velocity of rays in form of
\begin{eqnarray}
	\frac{{\rm{d}} T}{{\rm{d}} \lambda} & = & \frac{E}{f} \left( \sqrt{f_0} \mp
	\sigma \sqrt{f_0 - f} \sqrt{1 - \frac{f}{r^2} \left( \frac{E}{L} \right)^2}
	\right) \label{15}~,\\
	\frac{{\rm{d}} R}{{\rm{d}} \lambda} & = & \mp \frac{1}{C} \frac{E}{f} -
	\frac{\sigma \sqrt{f_0}}{C   f \sqrt{f_0 - f}} \sqrt{E^2 -
		\frac{L^2}{r^2} f} \label{16}~,\\
	\frac{{\rm{d}} \Phi}{{\rm{d}} \lambda} & = & \frac{L}{r^2} \label{17}~,
\end{eqnarray}
where $\mp$ represents in-falling or out-falling observers, respectively and $\sigma = \pm 1$ describes forwards or backward propagating rays, respectively. $E$ and $L$ are integral constants of null geodesic equations.

In observation, angular radius of shadow is determined by rays from the photon sphere. 
The integral constant $\frac{E}{L}$ of these rays can be obtained via $\frac{{\rm{d}} r}{{\rm{d}} \phi} = 0$, $\frac{{\rm{d}}^2r}{{\rm{d}} \phi^2} = 0$ and null geodesic equations. In the Schwarzschild-like space-time, the integral constant $\beta \equiv \left( \frac{E}{L} \right)_{ \rm{ps}} = \frac{f (r_{ \rm{ps}})}{r_{ \rm{ps}}^2}$. The location of photon sphere $r_{ \rm{ps}}$ is determined by $r f'(r) - 2 f(r) = 0$. As the integral constant $\beta$ indicates intrinsic property of light rays, the value of $\beta$ would remain the same in different reference frames.  

Using Eqs.~(\ref{16}), (\ref{17}) and value of $\beta$, we can obtain the angular radius $\psi_{ \rm{mov}}$
of shadow in generalized Lemaitre coordinate,
\begin{eqnarray}
	\cot \psi_{ \rm{mov}} & = & \sqrt{-\frac{g_{11}}{g_{33}}}\frac{{\rm d}R}{{\rm d }\Phi} = \frac{C \sqrt{f_0 - f} {\rm{d}} R}{r {\rm{d}}
		\Phi} \label{18a}\\
	& = &  \frac{r}{f} \left( \mp \beta \sqrt{f_0 - f} + \sigma \sqrt{f_0}
	\sqrt{\beta^2 - \frac{f}{r^2}} \right)~, \label{18}
\end{eqnarray}
where
\begin{equation}
	\sigma = \left\{\begin{array}{ll}
		1   & r_{ \rm{ps}} < r~,       \\
		- 1 & r_s < r < r_{ \rm{ps}}~.
	\end{array}\right.
\end{equation}
One may find that the angular radius $\psi_{ \rm{mov}}$ is independent of normalization constant $C$. In Appendix, this result would be compared with preview works.

\subsection{Aberration formulation}
Using the approach in section~\ref{II}, we would reconstruct aberration formulation in the Schwarzschild-like space-time in this subsection.

Firstly, we have already known how to calculate the angular radius $\psi_{ \rm{stat}}$ of the shadows for given spherical black hole \cite{synge_escape_1966},
\begin{equation}
\cot \psi_{ \rm{stat}} = \frac{\frac{1}{\sqrt{f}} {\rm{d}} r}{r {\rm{d}} \phi}
= \sigma \frac{r}{\sqrt{f}} \sqrt{\beta^2 - \frac{f}{r^2}}~.  \label{22}
\end{equation}
where the $\frac{{\rm d}r}{{\rm d}\phi}$ is determined by the light rays from photon sphere,
\begin{eqnarray}
\frac{{\rm{d}} t}{{\rm{d}} \lambda} & = & \frac{E}{f}\label{15-1}~,\\
\frac{{\rm{d}} r}{{\rm{d}} \lambda} & = &\frac{E \sigma}{\beta} \sqrt{\beta^2 - \frac{f}{r^2}} \label{16-1}~,\\
\frac{{\rm{d}} \phi}{{\rm{d}} \lambda} & = & \frac{E}{\beta r^2} \label{17-1}~,
\end{eqnarray}
The aberration formulation that describes the relation between $\psi_{\rm stat}$ and $\psi_{\rm mov}$ should be established by using Eqs~(\ref{18}) and (\ref{22}). 

Then, one might use coordinate velocities to describe relative motion between these reference frames and the speed of light. The 3-velocity of freely falling observers in the static coordinates (Eq.~(\ref{13-1})) can be given by
\begin{equation}
	\upsilon_o \equiv \frac{{\rm d} (r(\tau))}{{\rm d} (t(\tau))} =  \frac{u^1}{u^0}  =  \pm f \sqrt{1 - \frac{f}{f_0}}~. \label{20}
\end{equation}
From Eqs.~(\ref{16-1}) and (\ref{17-1}), the radial component of 3-velocity of light rays from photon sphere is of the form,
\begin{equation}
	c_r \equiv \frac{{\rm d} (r(\lambda))}{{\rm d} (t(\lambda))} =  \frac{\frac{{\rm{d}} r}{{\rm{d}} \lambda}}{\frac{{\rm{d}} t}{{\rm{d}}
			\lambda}} = \frac{f}{\beta} \sigma \sqrt{\beta^2 - \frac{f}{r^2}}~. \label{21}
\end{equation}
One should note that the covariant velocity of observers $u^{\mu}$ and null curve $\frac{{\rm{d}} x^{\mu}}{{\rm{d}} \lambda}$ are both described in static coordinate of spherical black hole. 
Using Eqs.~(\ref{20}), (\ref{21}) and (\ref{22}), we can express angular radius for freely
falling observers in term of that for static observers,
\begin{equation}
	\cot \psi_{ \rm{mov}} = \cot \psi_{ \rm{static}} \left( 1 -
	\frac{\upsilon_o}{c_r} \right) \sqrt{\frac{f_0}{f}} ~ \label{23}.
\end{equation}
In this case, one might find that the form of aberration formation is different from that obtained by Lebedev and Lake  \cite{lebedev_influence_2013} and reviewed by Perlick $et$ $al.$ \cite{perlick_black_2018}.

 

However, in the physical point of view, the observers receive the light rays from the photon sphere in the locally inertial frame. Usually, a set of orthogonal tetrads is introduced for calculating the black hole shadows. In this sense, the velocities should be expressed in the orthogonal tetrads. In the spherical black holes, we have
\begin{equation}
	(e^{(a)}_\mu) = 
	\left( \begin{array}{cccc}
	\sqrt{f} &   &   &  \\
	& \frac{1}{\sqrt{f}} &   & \\
	&   & r^2& \\
	&   &   & r^2 \sin^2\theta  \\
	\end{array}\right) ~, \label{21-1}
\end{equation}
From (\ref{13-1}) and (\ref{21-1}), the 4-velocities of observers in the orthogonal tetrads take the form of
\begin{equation}
	u^{(a)} = e^{(a)}_\mu u^\mu = \left(\sqrt\frac{f_0}{f},\pm \sqrt{ \frac{f_0}{f} -1}, 0, 0 \right)
\end{equation}
And from Eqs.~(\ref{15-1})-(\ref{17-1}) and (\ref{21-1}), the 4-velocities of the light rays from photon sphere in the orthogonal tetrads are
\begin{equation}
l^{(a)} \equiv e^{(a)}_\mu \frac{{\rm d}x^\mu}{{\rm d} \lambda} = 
\left(\frac{E}{\sqrt{f}}, \frac{E \sigma}{\beta} \sqrt{\frac{\beta^2}{f} - \frac{1}{r^2}} , \frac{E}{\beta r}, 0 \right)
\end{equation}
Then, we can obtain the radial component of these 3-velocities,
\begin{eqnarray}
	V_o \equiv \frac{u^{(1)}}{u^{(0)}} = \pm \sqrt{1-\frac{f}{f_0}} \label{24-1}\\
	C_r \equiv \frac{l^{(1)}}{l^{(0)}} = \frac{ \sigma}{\beta} \sqrt{\beta^2 - \frac{f}{r^2}} \label{25-1}
\end{eqnarray}
From Eqs~(\ref{18}) and (\ref{22}) and the expression of $V_o$ and $C_r$, the aberration formulation can take the form of
\begin{equation}
	\cot \psi_{ \rm{mov}} = \cot \psi_{ \rm{stat}} \left( 1 -
\frac{V_o}{C_r} \right) \frac{1}{\sqrt{1 - V_o^2}}
\end{equation}
In this case, one might find that the form of aberration formulation is the same as that in special relativity, Eq~(\ref{10}). This is consistent with the claims in previous works \cite{lebedev_influence_2013,grenzebach_aberrational_2015,perlick_black_2018,book:1533124}.

And in the case that light propagates
nearly along direction of $V_o$, we also have
\begin{equation}
\cot \psi_{ \rm{mov}} = \cot \psi_{ \rm{stat}} \sqrt{\frac{1 -
		V_o}{1 + V_o}}~.
\end{equation}

\subsection{Instances of angular radius of black holes shadow}

For the sake of intuitive, in this part, we consider parametrized Schwarzschild black hole \cite{tian_testing_2019} and Schwarzschild-de Sitter black hole for examples. They are representative cases of asymptotic and non-asymptotic space-time, respectively.

\subsubsection{Parametrized Schwarzschild black hole}

We consider the parametrized Schwarzschild black hole \cite{tian_testing_2019} with
\begin{equation}
	f (r) = \left( 1 - \frac{n   r_s}{r} \right)^{\frac{1}{n}} ~.
\end{equation}
Event horizon of the parametrized Schwarzschild black hole is located at $n 	r_s$. For $n = 1$, it returns to be Schwarzschild black hole. The $n$ deviated beyond 1 can be used to test modified  theory of gravity or exotic matters of black hole from observation. For example, authors of Ref.~\cite{tian_testing_2019} have presented a  solution in general relativity coupled to non-linear electrodynamics.

In space-time of a parametrized Schwarzschild black hole, the free-falling observers are always in-falling, which leads to $\upsilon_o < 0$. For Schwarzschild black hole $n=1$, we can obtain $\beta = \frac{2}{3 \sqrt{3} r_s}$ and $r_{ \rm{ps}} = \frac{3}{2} r_s$. The angular radius $\psi_{ \rm{mov}}$ of shadow with respect to freely-falling observers is of the form,
\begin{equation}
	\cot \psi_{ \rm{mov}} = r \sqrt{\frac{r (r_0 - r_s)}{r_0 (r - r_s)}}
	\left( \beta \sqrt{\frac{r}{r - r_s} - \frac{r_0}{r_0 - r_s}} + \sigma
	\sqrt{\frac{r \beta^2}{r - r_s} - \frac{1}{r^2}} \right) ~.
\end{equation}

\begin{figure}[ht]
	\centering
	{\includegraphics[width=1\linewidth]{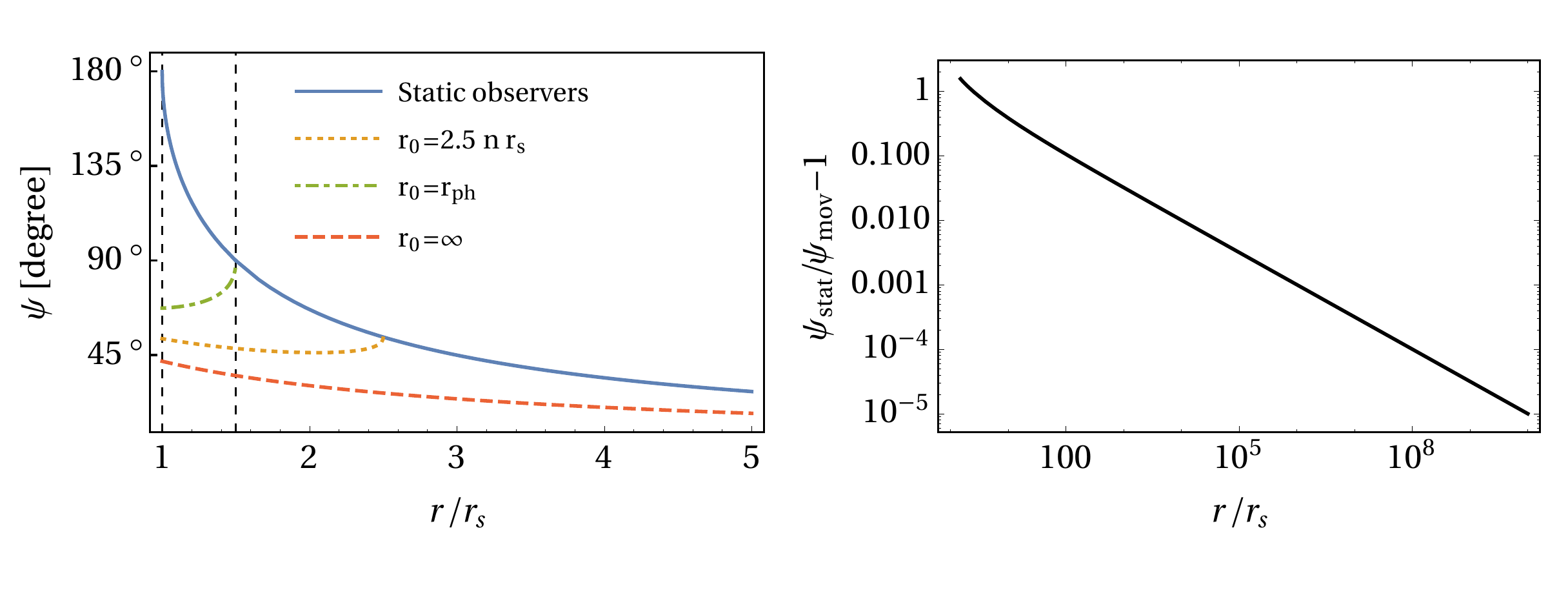}}
	\caption{Left panel: Angular radius as function of coordinate $r$ for Schwarzschild black hole. The solid line describes angular radius $\psi_{ \rm{staic}} (r)$ with respect to static observers. It's actually the results obtained by Synge \cite{synge_escape_1966}.  The others are angular radius $\psi_{ \rm{mov}}(r)$ observed by freely falling observers from spatial infinity, $2.5r_s$ and location of photon sphere, respectively. The freely falling observers undergo radial geodesic motion from solid line. Right panel: Relative deviation  between static and freely falling observers from spatial infinity as function of $r$.\label{Fig1}}
\end{figure}
In the left panel of Figure~\ref{Fig1}, we present angular radius as function of Schwarzschild coordinate $r$. For observers located at the large distance of black hole, the angular radius observed by static observers and freely falling observers are nearly the same. For the M87, we can estimate the distance, $d \approx 3 \times 10^{10} r_s$. It shows in the right panel of Figure~\ref{Fig1}  that there are relative deviation of angular radius around $10^{- 5}$ between static and freely falling observers. It seems beyond current accuracy of  measurement. On the other side, in the near region of black hole, there are two features. Firstly, in-falling observers' field of vision would not be filled with shadow of black hole, when they go through the photon sphere or event horizon. It suggests that freely observers in the interior of black hole can receive information of light from outside. Secondly, for freely falling observers  initially from finite distance, the angular radius  could decrease as they get closer to the black hole. The extreme case is that observers freely fall from photon sphere. They would always find the size of shadow decreasing as getting closer to the black hole.

\begin{figure}[ht]
	\centering
	{\includegraphics[width=1\linewidth]{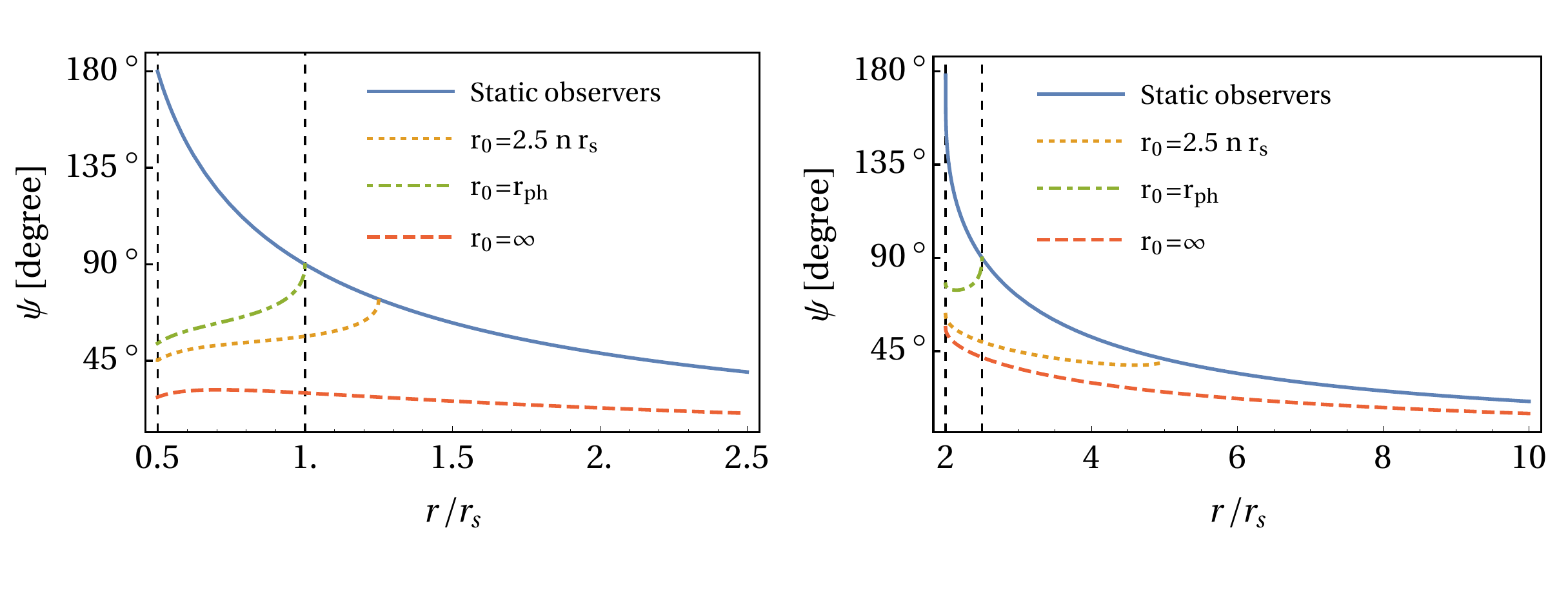}}
	\caption{Angular radius as function of coordinate $r$ for parametrized Schwarzschild black hole $n=0.5$ (left panel) and $n=2$ (right panel), respectively. The solid line describes angular radius $\psi_{ \rm{staic}} (r)$ with respect to static observers. The freely falling observers undergo radial geodesic motion from solid line. The dashed lines are angular radius $\psi_{ \rm{mov}}(r)$ observed by freely falling observers from spatial infinity, $2.5 n r_s$ and location of photon sphere, respectively. \label{Fig2}}
\end{figure}

For parametrized Schwarzschild black hole ($n \neq 1$), we can find that
location of photon sphere depend on parameter $n$,
\begin{equation}
	r_{ \rm{ps}} = \left( n + \frac{1}{2} \right) r_s~. \label{26}
\end{equation}
The integral constant of rays from photon sphere can be given by
\begin{equation}
	\beta_n = \frac{2 (2 n + 1)^{- \frac{1}{2 n} - 1}}{r_s}~. \label{27}
\end{equation}
Using Eqs.~ (\ref{18}), (\ref{26}) and (\ref{27}), we obtain angular radius of shadow with respect to
freely falling observers,
\begin{equation}
	\cot \psi_{ \rm{mov}} = r \left( \frac{1 - \frac{n   r_s}{r_0}}{1 -
		\frac{n   r_s}{r}} \right)^{\frac{1}{2 n}} \left( \beta_n
	\sqrt{\frac{1}{\left( 1 - \frac{n   r_s}{r} \right)^{\frac{1}{n}}} -
		\frac{1}{\left( 1 - \frac{n   r_s}{r_0} \right)^{\frac{1}{n}}}} +
	\sigma \sqrt{\frac{\beta^2_n}{\left( 1 - \frac{n   r_s}{r}
			\right)^{\frac{1}{n}}} - \frac{1}{r^2}} \right)~.
\end{equation}
In Figure~\ref{Fig2}, we present angular radius as function of coordinate $r$ for $n = 0.5$ and $2$. In the left panel of Figure~\ref{Fig2}, $n < 1$, angular radius of shadow tend to decrease when observers are closed to event horizon. We can calculate lower bound of angular radius  for  freely falling observers at event horizon,
\begin{eqnarray}
	\psi_{ \rm{mov},  \rm{sub}} & = &   \psi_{ \rm{mov}} |_{r_0 =
	\infty, r = n   r_s}\nonumber\\
	& = &  {\rm{arccot}} \left( \frac{(2 n + 1)^{\frac{1}{2 n} + 1}}{4 n} -
	\frac{n}{(2 n + 1)^{\frac{1}{2 n} + 1}} \right)~.
\end{eqnarray}
For Schwarzschild black hole, $n = 1$, $\cot \psi_{ \rm{mov},  \rm{sub}} =
	\frac{3 \sqrt{3}}{4} - \frac{1}{3 \sqrt{3}}$, which leads to
$\psi_{ \rm{mov},  \rm{sub}} \approx 42^{\circ}$.

In principle, we can throw a satellite into a black hole and analyse last observation of angular radius of black hole shadow. Observation of angular radius below the lower bound would suggest a modified theory of gravity or black hole made up of exotic matters.

\subsubsection{Schwarzschild-de Sitter black hole}

Likewise, we consider freely falling observer in Schwarzschild-de Sitter space-time with,
\begin{equation}
	f (r) = 1 - \frac{2 M}{r} - \frac{\Lambda}{3} r^2~. \label{30}
\end{equation}
There are two event horizons in the space-time on  condition of $\Lambda < 	(9 M^2)^{- 1}$. We symbolize locations of the horizons as $r_{H 1}$ and $r_{H 2}$, which satisfies $f (r_{H 1}) = f (r_{H 2}) = 0$ and $r_{H 1} < r_{H 2}$. We call them inner horizon $r_{H1}$ and cosmological horizon $r_{H2}$, respectively. In the near region, there is attractive gravity provided by black hole. In the large distance, the gravitational repulsion would domain space-time due to term of $\frac{\Lambda}{3}   r^2$. One can obtain the balance position between attractive and repulsive gravity, $r_{ \rm{bal}} = (3 M / 	\Lambda)^{\frac{1}{3}}$. For observers freely falling from location $r_0 < 	r_{ \rm{bal}}$, they would finally drop into black hole. For observers freely falling from location $r_0 > r_{ \rm{bal}}$, they would run away from the black hole. The latter can be understood as cosmological expansion perceived by co-moving observers.

From Eqs.~(\ref{18}) and (\ref{30}), we can obtain the angular radius with respect to freely falling observers,
\begin{eqnarray}
	\cot \psi_{ \rm{mov}} & = & \frac{r^2}{r - 2 M - \frac{\Lambda}{3} r^3}
	\left( \mp \beta \sqrt{2 M \left( \frac{1}{r} - \frac{1}{r_0} \right) +  \frac{\Lambda}{3} (r^2 - r_0^2)} \right. \nonumber \\
	& & \left.+ \sigma \sqrt{\left( 1 - \frac{2 M}{r_0} -
		\frac{\Lambda}{3} r_0^2 \right) \left( \beta^2 + \frac{\Lambda}{3} -
		\frac{1}{r^2} + \frac{2 M}{r^3} \right)} \right)~,
\end{eqnarray}
where $\beta^2 = \frac{1}{27 M^2} - \frac{\Lambda}{3}$. One should note the coordinate $r$ can be extended to infinity in the generalized Lemaitre coordinate for run-away observers.

\begin{figure}[ht]
	\centering
	{\includegraphics[width=.6\linewidth]{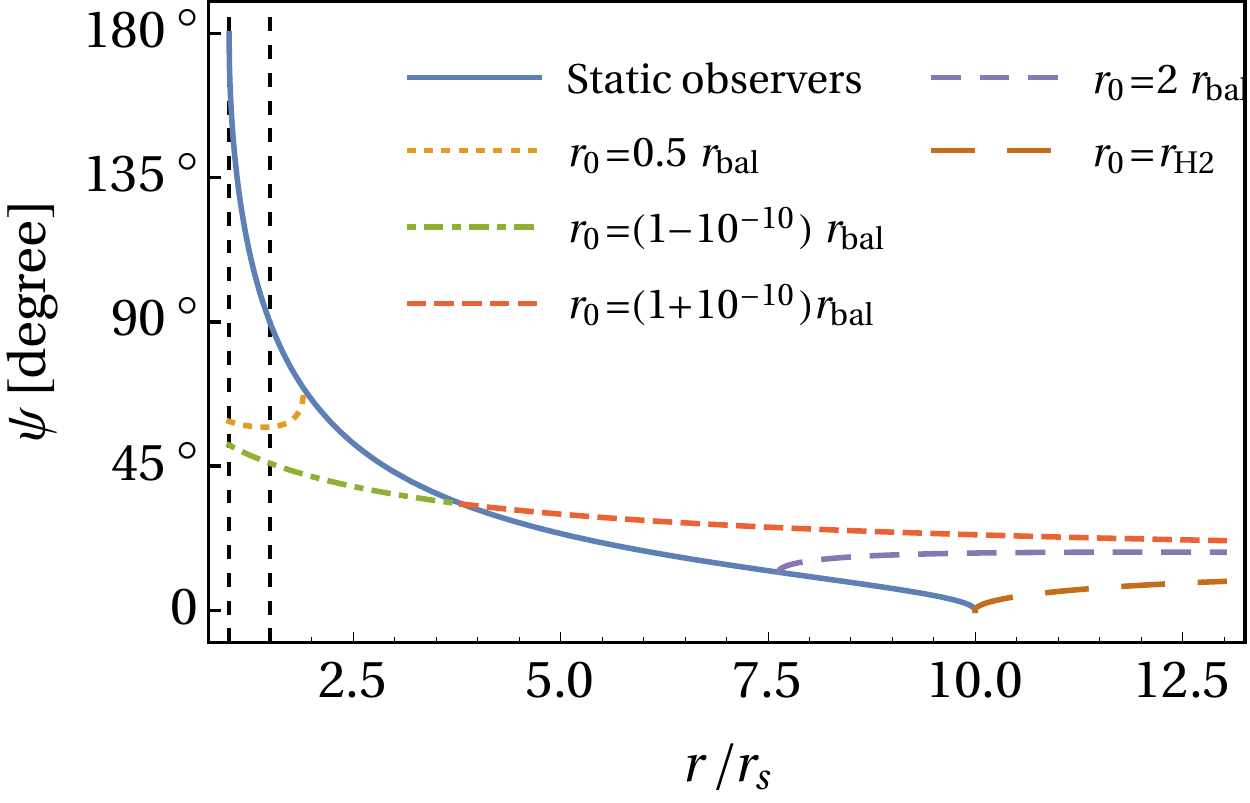}}
	\caption{Angular radius as function of coordinate $r$ for Schwarzschild-de Sitter black hole. The solid line describes angular radius $\psi_{ \rm{staic}} (r)$  with respect to static observers. The others are angular radius $\psi_{ \rm{mov}} (r)$ observed by freely falling observers from selected position. The freely falling observers undergo radial geodesic motion from the solid line.  Here, $r_s \equiv 2M$. And we set the cosmological constant $\Lambda=0.00675 M^{-2}$\label{Fig3}}
\end{figure}
In Figure~\ref{Fig3}, we present angular radius as function of Schwarzschild coordinate $r$. In the near region of black hole, especially $r_0 < r_{ \rm{bal}}$, the $\psi_{ \rm{mov}} (r)$ is closed to the description in Schwarzschild space-time. In the large distance of black hole, static and free-falling observers would perceive different angular radius of the shadow. The angular radius perceived by static observers tends to be vanished, while for the out-falling observer, angular radius tends to be a constant,
\begin{eqnarray}
	\cot\psi_{{\rm mov},\infty} \equiv \lim_{r \rightarrow \infty}\cot\psi_{\rm mov} & = & \lim_{r \rightarrow \infty} \frac{r}{f}\left(-\beta \sqrt{f_0 - f} + \sqrt{f_0}\sqrt{\beta^2 - \frac{f_0}{r^2}} \right) \nonumber\\
	& = & \lim_{r \rightarrow \infty} \frac{r}{-\frac{\Lambda}{3}r^2}\left(-\beta \sqrt{\frac{\Lambda}{3}r^2}\right) \nonumber\\
	& = & \beta \sqrt{\frac{3}{\Lambda}}~. \label{43}
\end{eqnarray}
 We expand expression of $\beta= \sqrt{\frac{1}{27M^2}-\frac{\Lambda}{3}}$ and rewrite Eq.~(\ref{43}) into sine,
\begin{eqnarray}
	\sin \psi_{ \rm{mov}, \infty} & = & \frac{1}{\sqrt{\cot^2 \psi_{{\rm mov},\infty} + 1}} \nonumber \\
	& = & \frac{1}{\sqrt{\left(\frac{1}{27M^2}-\frac{\Lambda}{3}\right){\frac{3}{\Lambda}}+1}} \nonumber\\
	& = & 3 M \sqrt{\Lambda}~.
\end{eqnarray}
This result is exactly the same as that obtained by Perlick~et~al.~\cite{perlick_black_2018}, in which they consider size of black holes shadow for a co-moving observer. It can be understood that the freely falling observers are actually co-moving with the expansion driven by ${\Lambda}$ at large distance. Besides, as it shown in Figure~\ref{Fig3}, there are two additional features. Firstly, for observers freely falling from the position beyond cosmological horizon $r_{H 2}$, namely, $r_0 > r_{H2}$, 
the calculation in the adapted coordinate can not give correct results. As in this case, $\cot\psi_{\rm mov}\sim\sqrt{f_0} \sim \sqrt{-1}$. We are not sure whether it suggests that those observers can not observe the shadow any more or it simply results from a breakdown of the coordinate system.
Secondly, we find that the angular radius of shadow could increase as freely falling observers run away from the black hole. By making use of equation ${\rm d}\psi_{\rm mov}/{\rm d}r = 0$ and its solution $r^*(>r_{\rm bal})$, we can conclude that for the observers out-falling from $r_H$, the size of the shadows would increase all the time. For the observers out-falling from the distance between $r_{\rm bal}$ and $r_H$, the size of the shadow would increase until the observers arrive at the position $r=r^*$.  If this is proved to be true, it can not be well understood without general relativistic effect that we shown here.

\section{Shadow-redshift relation}\label{IV}

We have shown the angular radius of shadow as function of radial coordinate $r$. The $r  $ is used to quantify the distance between black hole and observers, approximately. However, the coordinate $r$ usually is not observable. In this section, we would turn to relation between angular radius and gravitational redshift. In observation, if we can observe rays from photon sphere as it did by Event Horizon Telescope, the redshift of the photon sphere also seems not difficult to obtain.

Theoretically, we can calculate gravitational redshift of photon sphere for freely falling observers. From Eq.~(\ref{14}) and (\ref{15}), the redshift takes form of
\begin{eqnarray}
	1 + z_{\rm{mov}} & = & \frac{  - k^{\mu} u_{\mu} |_{ \rm{source}}}{-
	k^{\nu} u_{\nu} |_{ \rm{obs}}} \nonumber \\
	& = & \frac{\frac{f}{\sqrt{f (r_{ \rm{ph}})}}}{\sqrt{f_0} \mp \sigma
		\sqrt{f_0 - f} \sqrt{1 - \frac{f}{\beta^2 r^2}}}~,
\end{eqnarray}
where $k^\mu$ is wave vector of rays. Associating it with Eq.~(\ref{18}), we can obtain curve of angular  radius-gravitational redshift relation $(z(r), \psi(r))$ with parameter $r$. Likewise, in the following parts, we consider parametrized Schwarzschild and Schwarzschild-de Sitter black hole as representative examples.

\subsection{Parametrized Schwarzschild black hole}

In Schwarzschild space-time, the gravitational redshift from photon sphere observed by freely falling observers takes the form of
\begin{equation}
	1 + z_{ \rm{mov}} = \frac{\sqrt{\frac{3 r_0 (r - r_s)}{r (r_0 -
				r_s)}}}{\sqrt{\frac{r}{r - r_s}} + \sigma \sqrt{\frac{r_s (r_0 - r)}{r (r_0
				- r_s)}} \sqrt{\frac{r}{r - r_s} - \frac{27}{4} \frac{r_{_s}^2}{r^2}}} \label{34}~.
\end{equation}

In Figure~\ref{Fig4}, we present angular radius-redshift curve $(\psi, z)$ of freely falling observers. It has already distinguished with that observed by static
observers.
\begin{figure}[ht]
	\centering
	{\includegraphics[width=0.6\linewidth]{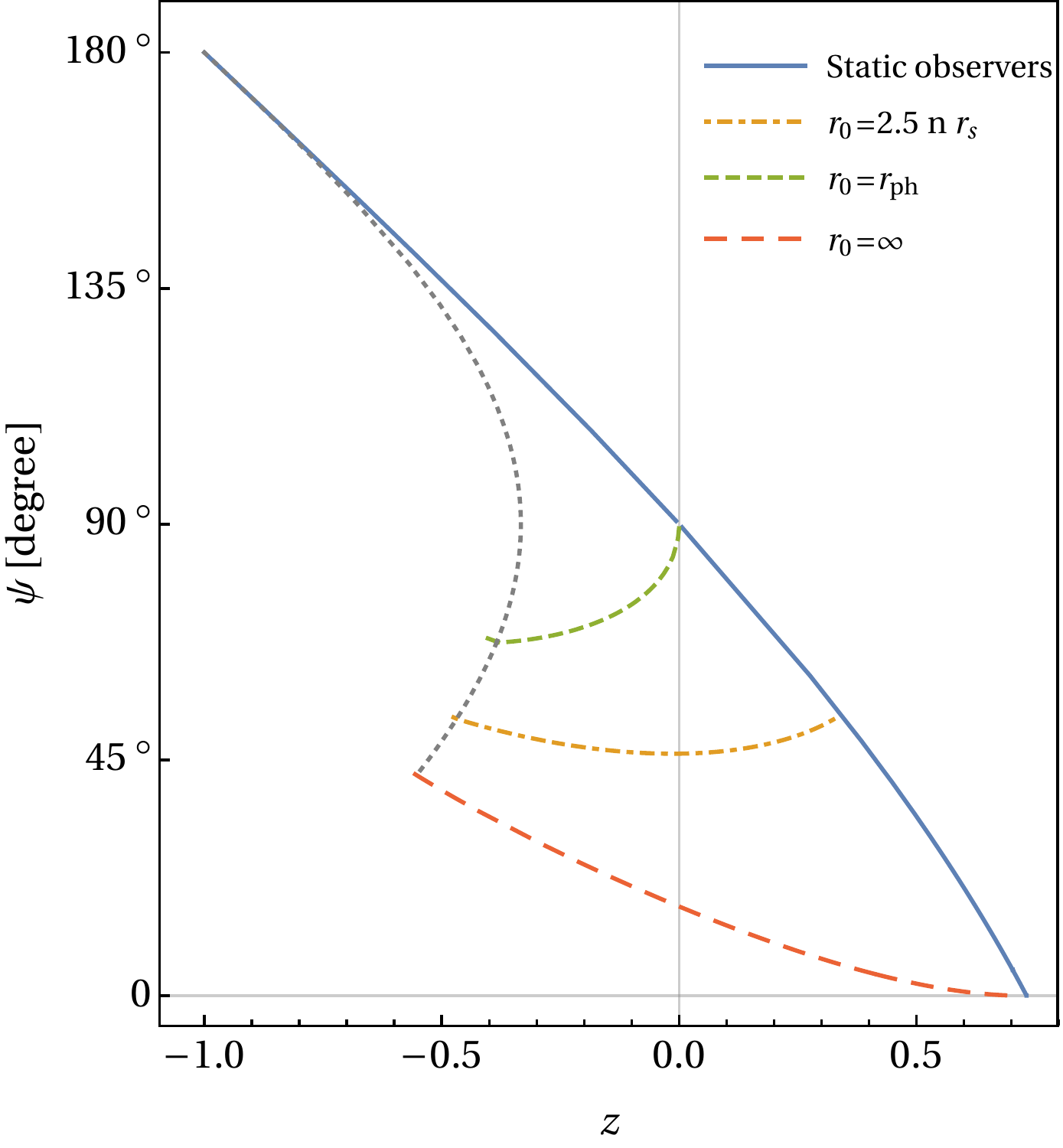}}
	\caption{Angular radius-gravitational redshift relations for Schwarzschild black hole. The solid line describes angular radius-gravitational redshift relation ($z_{\rm stat}(r)$, $\psi_{ \rm{staic}} (r)$) with respect to static observers. The other lines are angular radius-gravitational redshift relations ($z_{\rm mov}(r)$, $\psi_{ \rm{mov}} (r)$) with respect to observers  freely falling from selected initial position.\label{Fig4}}
\end{figure}
The dotted line is the curve of $(z_{ \rm{mov}}, \psi_{ \rm{mov}})_{r = r_s}$ for final position of freely falling observers. The in-falling observers would start from positions on solid line and finally reach the dotted line when they pass through event horizon. Here, we wish to point out two features. Firstly, as it shown via the dotted line, all the observers would perceive blueshift at the horizon. And we can calculate the gravitational redshift at horizon from
Eq.~(\ref{34}),
\begin{equation}
	z_{ \rm{mov}} |_{r = r_s} = \frac{4}{3} \left( \frac{2}{3
		\sqrt{3}} \frac{1}{\sqrt{1 - \frac{r_s}{r_0}}} + \frac{3 \sqrt{3}}{2}
	\sqrt{1 - \frac{r_s}{r_0}} \right)^{- 1} - 1~.
\end{equation}
It indicates that blueshift received by freely falling observers is larger than $- 1$. The extreme case is that observers are static at event horizon and observers freely fall from positions closed to event horizon. They can receive a photon with infinity energy from the photon sphere. It seems a classical firewall that would burn up anything stayed at the horizon. Secondly, for freely falling observers, angular radius-redshift relation is not always monotonic. Some of them could find angular radius decreasing as the redshift get lower. 

Furthermore, we consider parametrized Schwarzschild black hole, ($n \neq 1$). The gravitational redshift with respect to freely falling observers takes form of
\begin{equation}
	1 + z_{ \rm{mov}} = (2 n + 1)^{\frac{1}{2 n}} \left( \frac{1 - \frac{n
			r_s}{r}}{1 - \frac{n   r_s}{r_0}} \right)^{\frac{1}{2 n}}
	\left( \left( 1 - \frac{n   r_s}{r} \right)^{- \frac{1}{2 n}} \mp
	\sigma \sqrt{1 - \left( \frac{1 - \frac{n   r_s}{r}}{1 - \frac{n
				r_s}{r_0}} \right)^{\frac{1}{n}}} \sqrt{\left( \frac{1}{1 -
			\frac{n   r_s}{r}} \right)^{\frac{1}{n}} - \frac{1}{\beta^2 r^2}}
	\right)^{- 1}~.
\end{equation}
In Figure~\ref{Fig5}, we show angular radius-redshift relation $(z, \psi)$ for $n = 0.5$ and $n = 2$. Smaller $n$ tends to indicate a larger blueshift observed by the in-falling observers at horizon. 
For $n = 0.5,$ the angular radius of shadow tend to decrease as the redshift get lower. Conversely, for $n = 2$, angular radius of shadow tend to increase as the redshift get lower.

\begin{figure}[ht]
	\centering
	{\includegraphics[width=0.9\linewidth]{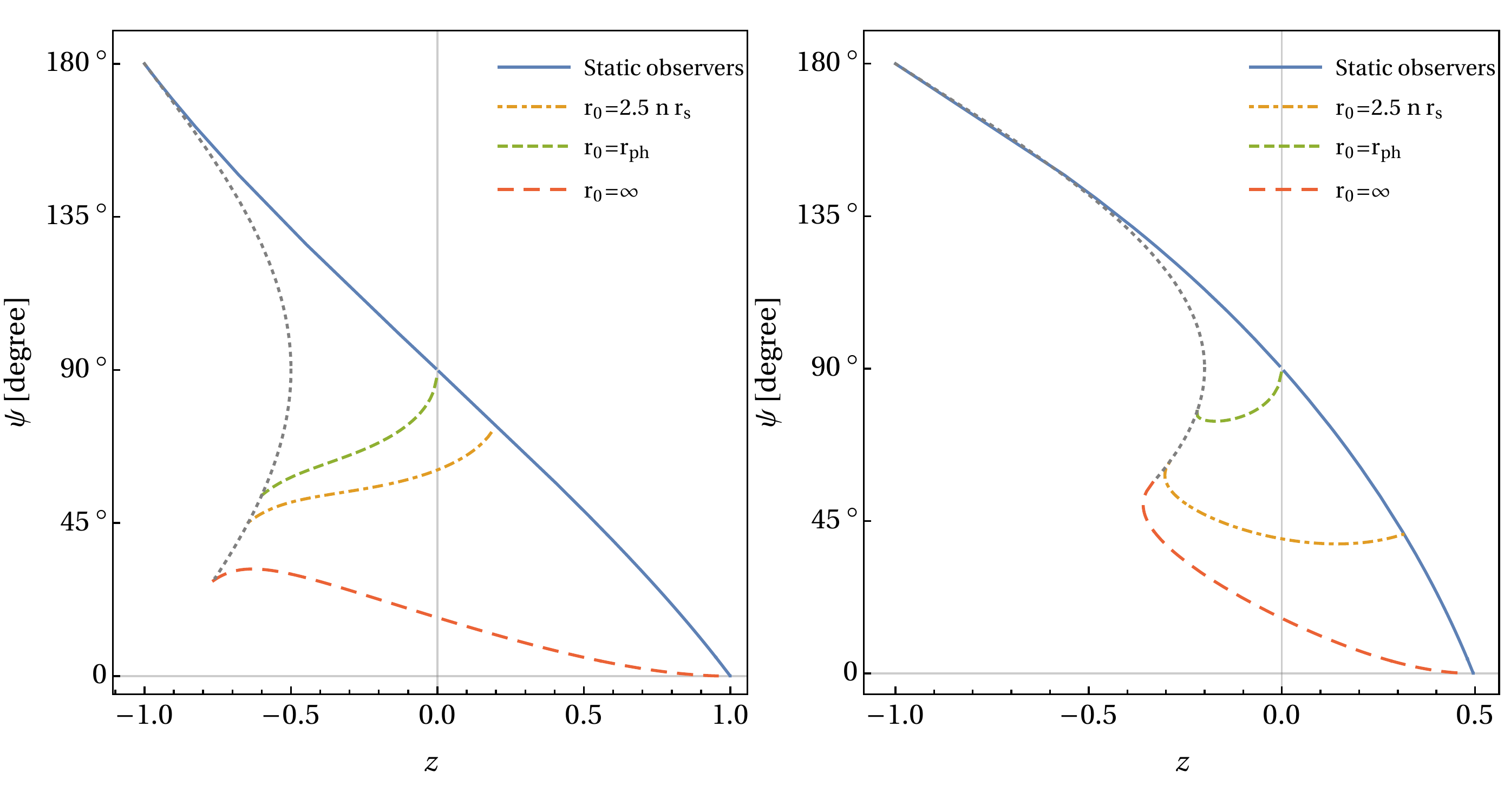}}
	\caption{Angular radius-gravitational redshift relations for parametrized Schwarzschild black hole. Left panel and right panel are curves  with parameters $n=0.5$ and $2$, respectively.  The solid line describes angular radius-gravitational redshift relation ($z_{\rm stat}(r)$, $\psi_{ \rm{staic}} (r)$) with respect to static observers. The other lines are angular radius-gravitational redshift relations ($z_{\rm mov}(r)$, $\psi_{ \rm{mov}} (r)$) with respect to observers  freely falling from selected initial position. \label{Fig5}}
\end{figure}

\subsection{Schwarzschild-de Sitter black hole}

Likewise, we consider angular radius-redshift relation with respect to
freely falling observers in Schwarzschild-de Sitter space-time. From
Eq.~(\ref{34}), the gravitational redshift takes the form of
\begin{equation}1 + z_{ \rm{mov}} = \frac{\left( 1 - \frac{2 M}{r} - \frac{\Lambda}{3}
		r^2 \right) \sqrt{\frac{3}{1 - 9 M^2 \Lambda}}}{\sqrt{1 - \frac{2 M}{r_0} -
			\frac{\Lambda}{3} r^2_0} \mp \frac{\sigma}{\beta} \sqrt{2 M \left(
			\frac{1}{r} - \frac{1}{r_0} \right) + \frac{\Lambda}{3} (r^2 - r_0^2)
			\left( \beta^2 + \frac{\Lambda}{3} - \frac{1}{r^2} + \frac{2 M}{r^3}
			\right)}}~, \end{equation}
where $\beta = \sqrt{\frac{1}{27 M^2} - \frac{\Lambda}{3}}$. When observers are far
away from the black hole, we can obtain
\begin{equation}
	\lim_{r \rightarrow \infty} \frac{z_{ \rm{mov}}}{r} = \sqrt{\Lambda}~.
\end{equation}
It's similar to the Hubble's law with Hubble constant $\tilde{H}_0 \equiv 	\sqrt{\Lambda}$. Thus, we would use $\sqrt{\Lambda}$ to estimate how the expansion rate of the current universe  affects angular radius of shadow. It's kind of reasonable, as current universe is in the epoch of dark energy dominated. But, strictly speaking, the cosmological constant in Schwarzschild de-Sitter space-time has no relevance to cosmology, which is described by FLRW space-time.

In Figure~\ref{Fig6}, we present angular radius-redshift curve $(\psi, z)$ for freely falling observers in Schwarzschild de-Sitter space-time. The solid line is the angular radius-redshift curve with respect to static observers. In the near region of black hole, $r_0 < r_{ \rm{bal}}$, the angular radius-redshift relation $(\psi, z)$ is closed to that sketched in Schwarzschild space-time. For observers initially located far away from the balance position $r_{ \rm{bal}}$, namely, $r_0 > r_{ \rm{bal}}$, the angular radius could increase or decrease with redshift. The extreme case is that observers initially locate at horizon $r_{H 2}$. They can observe the black hole shadow emerged in their field of vision until the angular radius turns to be a constant.

With the first M87 Event Horizon Telescope results that angular radius $\psi \approx 4 \mu  \rm{as}$ and redshift $z \approx 0.0039$, we can consider redshift and size of shadow with a realistic cosmological constant $\Lambda \sim 10^{- 52}$ m$^{- 2}$ and black hole mass $M \approx 6.5 \times 10^9 M_{\odot}$ in the Schwarzschild-de Sitter space-time. In Figure~\ref{Fig7}, we present gravitational redshift $z_{\rm mov}$ of light rays from photon sphere for freely falling observers as function of coordinate $r$. In the left panel of Figure~\ref{Fig7}, it shows that the redshift is dominated by central black hole $M$ in near region and is dominated by cosmological constant $\Lambda$ in large distance. It suggests that we can extract part of gravitational redshift $\Delta z_{\rm gra} \equiv z_{\rm mov}(r= r_{\rm bal})$ from $z_{\rm mov}$, which leads to an effective cosmological redshift $z_{\rm cos}$ in large distance,
\begin{equation}
	1 + z_{\rm cos} \equiv \frac{1 + z_{\rm mov}}{1+ \Delta z_{\rm gra}}~.
\end{equation}
In observation, $z_{\rm cos}$ corresponds to observed redshift of M87. Thus, we could plot M87 with observation data \cite{collaboration_first_2019} in Figure~\ref{Fig7}. It shows that our calculation is closed to the observation of M87. In the right panel of Figure~\ref{Fig7}, we present effective cosmological redshift $z_{\rm cos}$ as function of coordinate $r$. In large distance, one can find that the effective cosmological redshift with respect to freely falling observers would take form of
\begin{equation}
	1 + z_{\rm cos} \rightarrow \frac{1+ \Delta z_{\rm gra} + \sqrt{\Lambda}(r-r_{\rm bal})}{1+ \Delta z_{\rm gra}}~.
\end{equation}

In Figure~\ref{Fig8}, we present angular radius-gravitational redshift relation for selected observers. The cosmological constant and black hole mass are the same as those used in Figure~\ref{Fig7}. In the right panel of Figure~\ref{Fig8}, we focus on freely falling observers in large distance and present angular radius as function of effective cosmological redshift $z_{\rm cos}$. For M87, we can predict that the size of shadow would decrease as expansion of the universe. Besides, for angular radius less than $10^{-12}$rad, there seems a rather rich structure of angular radius-redshift relation. Firstly, angular radius of shadow tends to a constant $3M\sqrt{\Lambda}$ as redshift increases. Secondly, there exist black holes whose angular radius of shadow would increase with redshift. Thirdly, black holes with smaller size of shadow have more possibility to observe blueshift. As measurement precision improves in the future, it's possible to test these structure of angular radius-redshift relation in observation. 

\begin{figure}[ht!]
	\centering
	{\includegraphics[width=0.6\linewidth]{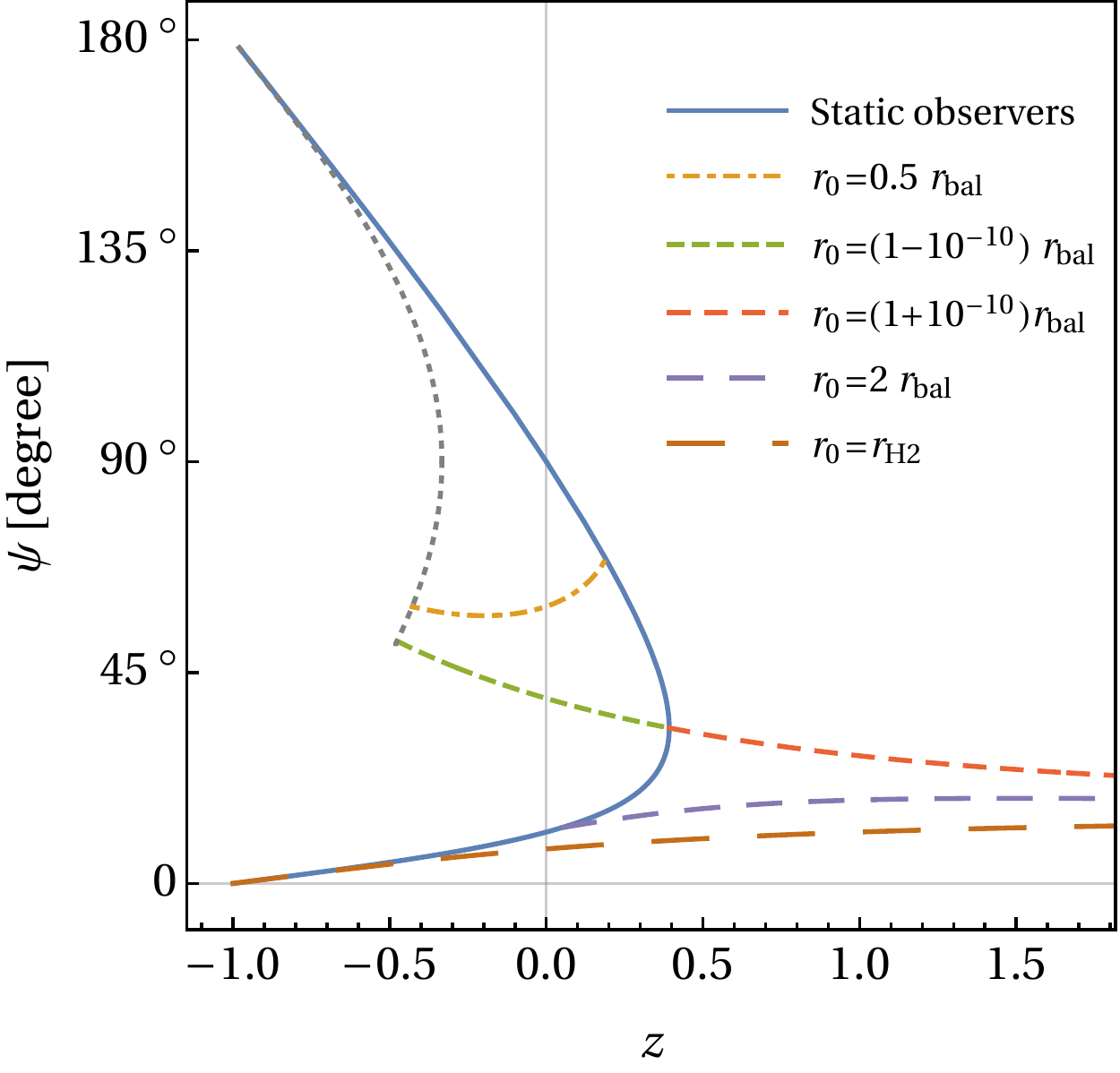}}
	\caption{Angular radius-gravitational redshift relations for Schwarzschild-de Sitter black hole. The solid line describes angular radius-gravitational redshift relation ($z_{\rm stat}(r)$, $\psi_{ \rm{staic}} (r)$) with respect to static observers. The others are angular radius-gravitational redshift relations ($z_{\rm mov}(r)$, $\psi_{ \rm{mov}} (r)$) with respect to observers  freely falling from selected positions. The cosmological constant is set $0.00675M^{-2}$. \label{Fig6}}
\end{figure}

\begin{figure}[ht!]
	\centering
	{\includegraphics[width=.95\linewidth]{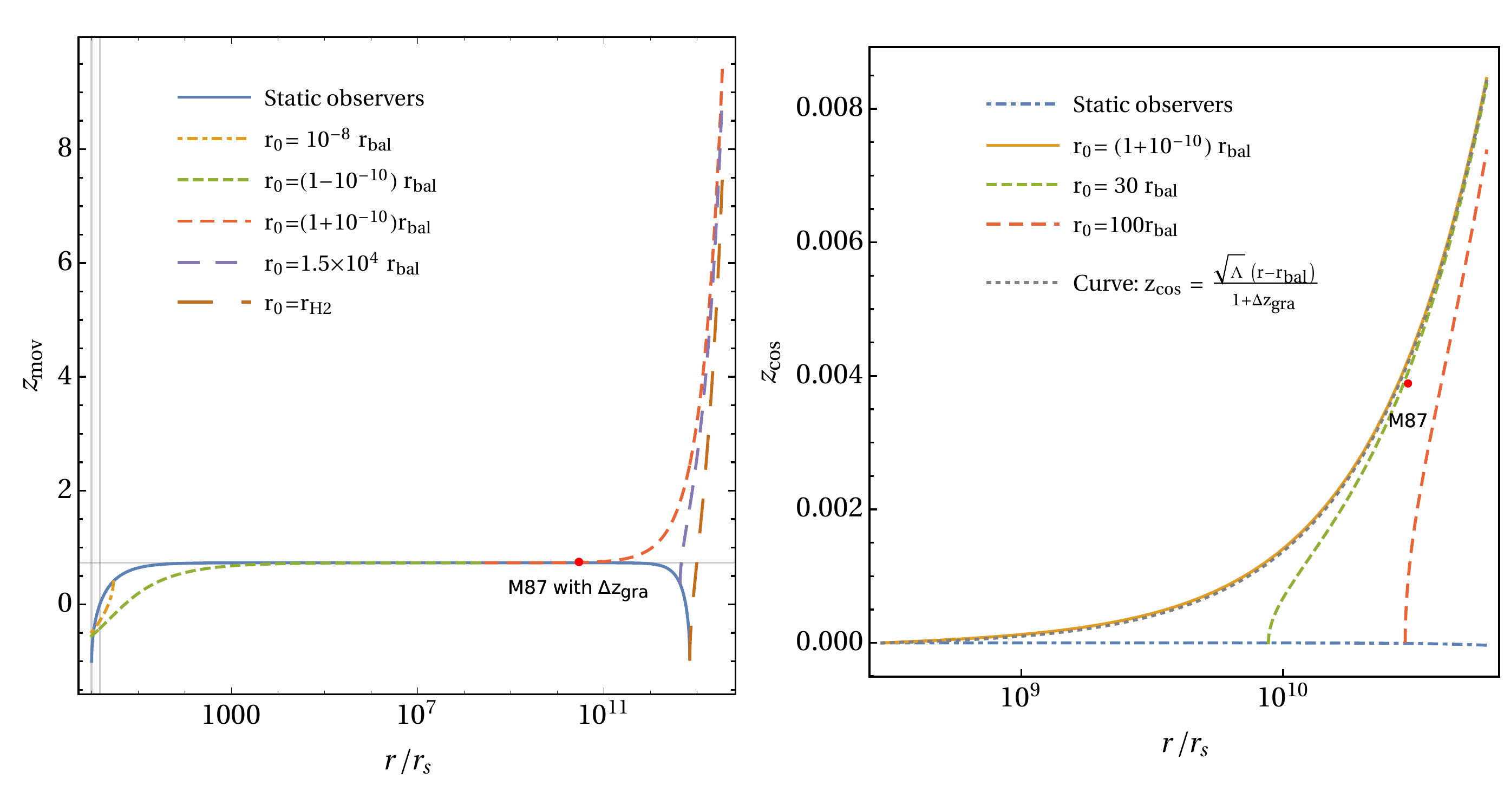}}
	\caption{Left panel: Gravitational redshift of rays from photon sphere as function of Schwarzschild coordinate $r$ for selected observers. The observers are static in the space or freely fall from an initial position $r_0$, respectively. Right panel: Effective cosmological redshift as function of Schwarzschild coordinate $r$ with respect to selected observers. In both plots, we plot M87 with observation data \cite{collaboration_first_2019}. In the Schwarzschild-de Sitter space-time, mass of black hole is set $6.5 \times 10^9 M_{\odot}$ and cosmological constant is set $10^{-52} {\rm m^{-2}}$} \label{Fig7}
\end{figure}

\begin{figure}[ht!]
	\centering
	{\includegraphics[width=.95\linewidth]{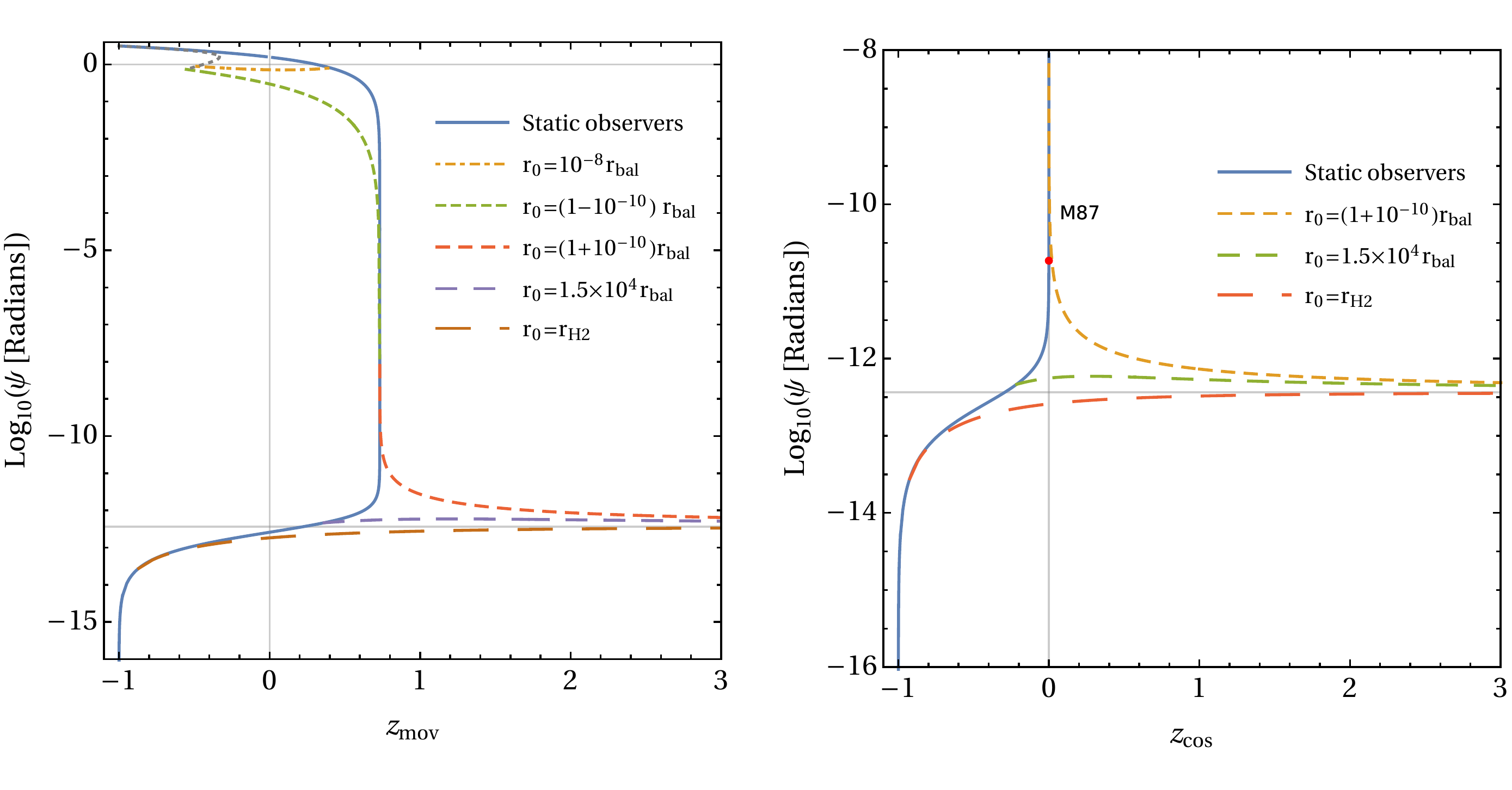}}
	\caption{Left panel: Angular radius-gravitational redshift relation for selected observers. The solid line describes angular radius-gravitational redshift relation ($z_{\rm stat}(r)$, $\psi_{ \rm{staic}} (r)$) with respect to static observers. The others are angular radius-gravitational redshift relations ($z_{\rm mov}(r)$, $\psi_{ \rm{mov}} (r)$) with respect to observers  freely falling from selected positions.  Right panel: Angular radius-effective cosmological redshift relation for large distant observers. Here, we also plot M87 with observation data \cite{collaboration_first_2019}. Both of plots are considered in Schwarzschild-de Sitter space-time, in which mass of black hole is set $6.5 \times 10^9 M_{\odot}$ and cosmological constant is set $10^{-52} {\rm m^{-2}}$.\label{Fig8}}
\end{figure}

\section{Black hole shadow in the view of freely falling observers} \label{V}

In adapted coordinate,  coordinate time $T$ is well-defined and describes the physical time that a freely falling observer perceive. If someone in a spacecraft approaching a black hole, the simplest scientific measurement that he or she can do is recording the size of shadow in the view at every moment. This suggests that expressing physical observables with explicit coordinates $(T,R,\Theta,\Phi)$ in the moving reference frame is meaningful. 

From Eqs.~(\ref{14}), we consider a freely falling observer in the adapted coordinate,
\begin{eqnarray}
T      & = & \sqrt{f_0} (t - t_0) \mp \int^r_{r_0} {\rm{d}} r \left\{
\frac{\sqrt{f_0 - f}}{f} \right\}~,                                         \label{35} \\
R_0      & = & \mp \sqrt{f_0-1}  (t - t_0) + \sqrt{f_0-1} \int_{r_0}^r {\rm{d}} r
\left\{ \frac{1}{f} \sqrt{\frac{f_0}{f_0 - f}} \right\}~,  \label{36}     
\end{eqnarray} 
where $R=R_0$, $\Theta$ and $\Phi$ are constants. From Eqs.~(\ref{35}) and (\ref{36}), we can express coordinate time $T$ in terms of $r$,
\begin{equation}
T = \mp \left( \frac{R_0}{\sqrt{1-\frac{1}{f_0}}} - \int_{r_0}^{r}\frac{{\rm d}r}{\sqrt{f_0 - f}} \right)~, \label{37}
\end{equation}
where $\pm$ represent out-falling or in-falling observers, respectively. The coordinate time $T$ describes physical time of the freely falling observers. Substituting $r$ with $T$ for Eq.~({\ref{18}}), we can study  angular radius of shadow evolving with time, $\psi_{\rm mov}(T,R_0)$. It's rather physical, as in the view of freely falling observers, what they actually perceive is the size of shadow changing with time. 

Specifically, from Eq.~(\ref{37}) for $R_0=0$, coordinate time $T$ of the observers is
\begin{equation}
	T = \pm  \int_{r_0}^{r}\frac{{\rm d}r}{\sqrt{f_0 - f}}~. \label{56}
\end{equation}
Using Eq.~(\ref{56}), we can express angular radius in term of coordinate time $T$,
\begin{equation}
	\psi_{\rm mov}(T,0) = \cot^{-1}\left(\cot\psi_{\rm stat}(r(T,0))\left(1-\frac{V_o(r(T,0))}{C_r(r(T,0))}\right)\frac{1}{\sqrt{1-V_o^2(r(T,0))}}\right)~.
\end{equation}
Then, we can study the time evolution of $\psi_{\rm mov}(T,0)$ via plotting.

In the left panel of Figure \ref{Fig9}, we present angular radius of shadow in Schwarzschild space-time for in-falling observers. All of them could observe rays from photon sphere when they go through inner horizon of black hole. And they would find size of shadow decreasing with time at the beginning when they undergo freely falling motion. In the right panel of Figure \ref{Fig9}, we present angular radius of shadow in Schwarzschild-de Sitter space-time for out-falling observers. In the view of the observers, some of them can perceive shadow of black hole getting larger as time goes by. After a long time, all of them would find the size of shadow approaching a constant.
\begin{figure}[ht!]
	\centering
	{\includegraphics[width=1\linewidth]{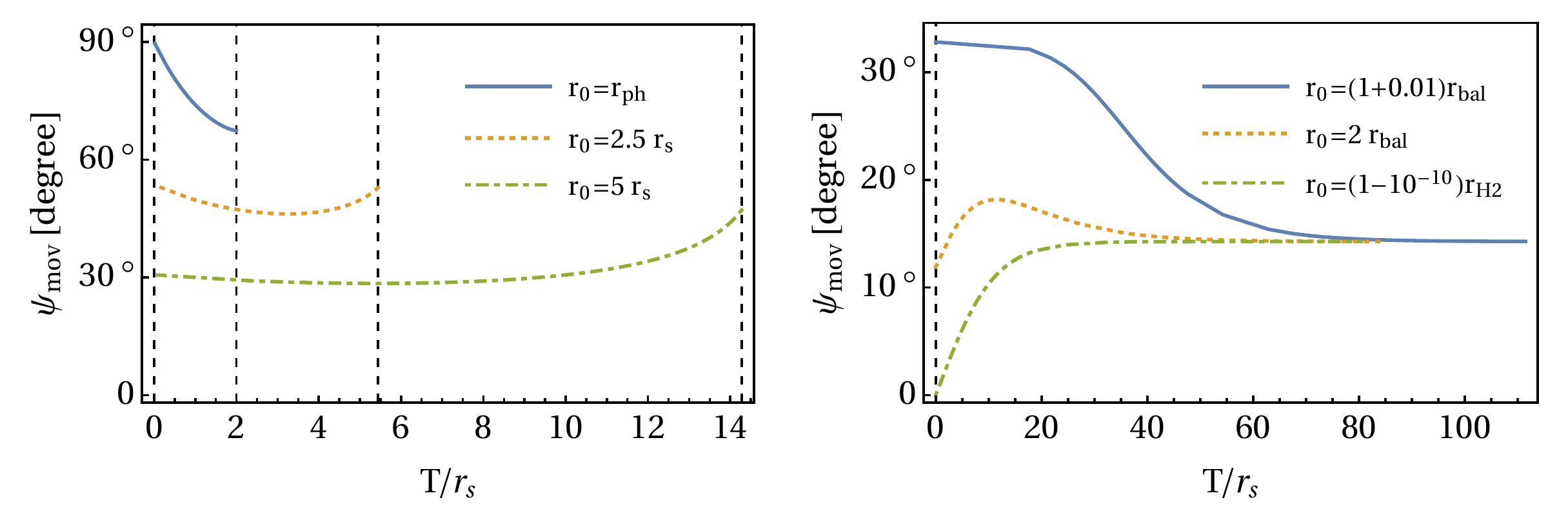}}
	\caption{Left panel: Angular radius $\psi_{\rm mov}(T,0)$ as function of coordinate time $T$ in Schwarzschild space-time for freely in-falling observers. The vertical lines represent the time at which the observers reach inner horizon of black hole. Right panel: Angular radius $\psi_{\rm mov}(T,0)$ as function of coordinate time $T$ in Schwarzschild-de Sitter space-time for freely out-falling observers. Here, $r_s=2M$ and cosmological constant $\Lambda = 0.00675M^{-2}$. In both plots, the observers are initially located at $r_0$. \label{Fig9}}
\end{figure}

\section{Conclusions and Discussions}\label{VI}

In this paper, we focused on angular radius of black hole shadow with respect to freely falling observers. The freely falling reference frame was described by generalized Lemaitre coordinate in Schwarzschild-like spherical space-time. In this coordinate, we derived aberration formulation and angular radius-redshift relation. For the sake of intuitive, we considered parametrized Schwarzschild black hole and Schwarzschild-de Sitter black hole as representative examples. We found that the freely falling observers would observe finite size of shadow, when they go through inner horizon. It suggests that the freely observers in the interior of black hole could receive information of light from outside. For Schwarzschild black hole, a freely falling observer from spatial infinity would observe the angular radius of shadow at horizon,$\psi_{{ \rm{mov}},r_s} \approx 42^{\circ}$. It's different from the size of angular radius with respect to static observers \cite{synge_escape_1966}. The static observers would observe nothing at horizon. For Schwarzschild-de Sitter black hole, the space-time is not asymptotic. We found the angular radius of shadow could increase at the beginning when observes are moving farther from the black hole.

Perlick, Tsupko and Bisnovatyi-Kogan \cite{perlick_black_2018} firstly  considered black hole shadow with respect to co-moving observers of the expanding universe. With Schwarzschild-de Sitter space-time, they found that the angular radius of shadow tend to be a constant $\sin \psi_{ \rm{mov}, \infty} = 3 M \sqrt{\Lambda}$ at large distance. In our approach, we  obtained consistent results of the angular radius. It's because that the freely falling observers are actually co-moving with the expansion driven by ${\Lambda}$ at large distance.

Besides, we also found something anormal. The calculation shown that increasing angular radius does not necessarily mean that we are getting closer to black hole and vice versa. It might indicate that motional status of reference frame could affect observation more significantly than expected in
the space. 

For the first M87 Event Horizon Telescope results, we can show that our calculation is closed to the observation of M87. Besides, as it shown in the right panel of Figure~\ref{Fig8},  there seems a rather rich structure of angular radius-redshift relation for those black holes with angular radius less than 0.2  $\rm \mu as$ and mass around $6.5 \times 10^9 M_{\odot}$. Perhaps, as measurement precision improves in the future, it might be well-motivated to test these structure in observation.

\acknowledgments
	The authors wish to thank Prof.~Sai Wang for discussions and thank anonymous referee(s) for valuable comments. One of us (Q.-H. Zhu) thank Jie Jiang for introducing the topic of black hole shadow. This work has been funded by the National Nature Science Foundation of China  under grant No. 11675182 and 11690022.

\appendix
\section{Remarks on calculation of angular radius of the shadows}
In the previous section, we calculated angular radius in Lemaitre coordinate and reconstructed the aberration formulation for freely falling observers. In the reference frame, we simply used formulation of angular radius Eq.~(\ref{18a}) proposed by Synge \cite{synge_escape_1966}, which suggests
\begin{equation}
\cot\psi= \sqrt{\frac{g_{11}}{g_{33}}}\frac{k^1}{k^3}~, \label{25-1}
\end{equation}
where $k^{\mu}$ is 4-velocity of light ray. As Synge \cite{synge_escape_1966} only consider static coordinate of Schwarzschild space-time, one might wonder whether it's reasonable for the  freely falling reference we considered here.  In the following, we would clarify this via showing that the formulation (Eqs.~(\ref{18a}) or (\ref{25-1})), in fact, can be derived from standard astrometric observable \cite{Soffel:2019aoq} in the case of spherical space-time.

In astrometry, what we can observe is angle between two incident light rays. It's formulated as
\begin{equation}
\cos \psi = \frac{\gamma^*w \cdot \gamma^*k}{|\gamma^*w||\gamma^*k|}~, \label{25}
\end{equation}
where $w$ and $k$ are null vectors that represent two incident rays,  $\gamma^*$ is projector for given observers $u$, namely, $\gamma^\mu_\nu=\delta^\mu_\nu + u^\mu u_\nu$, and inner product is defined in term of metric $g_{\mu \nu}$. Expanding the projector, we can rewrite the Eq.~(\ref{25}) as
\begin{equation}
\cos \psi = \frac{ k \cdot w}{(u \cdot k)(u \cdot w)} +1~. \label{26-1}
\end{equation}
In order to compare it with Synge's proposal (Eq.~(\ref{25-1})), we rewrite Eq.~(\ref{26-1}) in term of cotangent,
\begin{equation}
\cot \psi = {\rm sign}\left(\frac{\pi}{2} - \psi \right)\sqrt{-1-\left(\frac{1}{k \cdot w}\right) \frac{(u \cdot k)^2 (u \cdot w)^2}{k \cdot w + 2 (u \cdot k)(u \cdot w)}}~.
\end{equation}
\begin{figure}[ht]
	\centering
	{\includegraphics[width=0.7\linewidth]{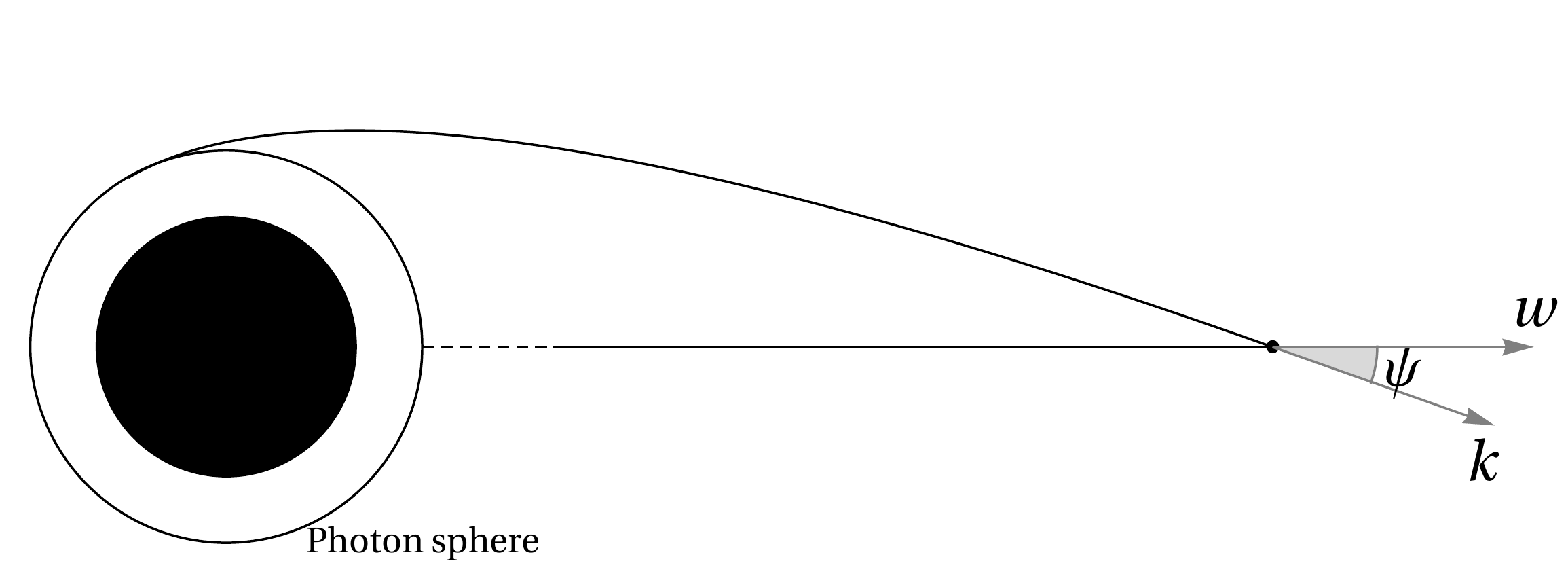}}
	\caption{Schematic diagram of spherical black hole shadow in astrometry. The $k$ and $w$ are a bending light ray from photon sphere and an arbitrary radial light ray, respectively. The angle $\psi$ is observed angular radius of black hole shadow. \label{Fig0}}
\end{figure}
The schematic diagram for spherical black hole shadow in astrometry is shown in Figure \ref{Fig0}. What we observe is the angle $\psi$ between a radial ray $w$ and a bending ray $k$ from photon sphere. Here, the angle $\psi$ represents the observed angular radius of black hole shadow. Because of spherical symmetry of space-time, the metric is diagonal and we can consider $\theta=\frac{\pi}{2}$ for simplicity. In adapted coordinate, we have $u^{\mu}=(-g_{00})^{1/2}\delta^{\mu}_0$. Associated with $k \cdot k = w \cdot w =0$, it's straightforward to get the following expression,
\begin{eqnarray}
\cot \psi & = & {\rm sign}\left(\frac{\pi}{2} - \psi \right)\sqrt{-1-\left(\frac{1}{g_{00}k^0w^0+g_{11}k^1w^1}\right)\frac{(g_{00}k^0w^0)^2}{g_{00}k^0w^0+g_{11}k^1w^1-2g_{00}k^0w^t}} \nonumber \\
& = & {\rm sign}\left(\frac{\pi}{2} - \psi \right) \sqrt{\frac{g_{11}}{g_{33}}}\left|\frac{k^1}{k^3}\right| \nonumber \\
& = & \sqrt{\frac{g_{11}}{g_{33}}}\frac{k^1}{k^3}~. \label{29}
\end{eqnarray}
In third equal, the sign is determined since $\psi$ is in the range of $[0,\pi]$. It indicates that the Synge's proposal (Eq.~(\ref{25-1})) is consistent with the definition of observed angle in astrometry (Eq.~(\ref{25})). And the Eq.~(\ref{18a}) is reasonable for freely falling reference frame. 

From astrometry of shadow shown in Figure~\ref{Fig0}, one might criticize that  derivation of angular radius $\psi$ is unphysical, because we could not observe radial light rays from centre of black hole. Fortunately, it can be solved via measurement of angular diameter of shadow. Namely, one can consider two bending light rays from photon sphere in a plane. And the observed angular diameter is exactly twice of $\psi$. It indicates that $\psi$ describes size of spherical black hole shadow well.

\section{Compare Eq.~(\ref{18}) with previous works}
Besides angular radius calculated in adapted coordinate for moving observers with Synge's proposal (Eq.~(\ref{25-1})), there is other approach. Lebedev and Lake \cite{lebedev_influence_2013,lebedev_relativistic_2016} calculated angular distance via inserting 4-velocity of observers into Eq.~(\ref{26-1}). Their results also can be used to calculate angular radius of black hole shadow. As the two approaches are both derived from Eq.~(\ref{25}), the results of these approaches should be, at least, numerically consistent. However, the angular radius $\alpha_{\rm{radial}}$ given by Lebedev and Lake for radial freely falling observers is in a different form from Eq.~(\ref{18}) in this paper. Thus, here, we would check the equivalence, directly,
\begin{eqnarray}
\cos \psi_{\rm{mov}} & = & \frac{\cot \psi_{\rm mov}}{\sqrt{1 + \cot^2
		\psi_{\rm{mov}}}} \nonumber\\
& = & \frac{\frac{r}{f} \left( - \beta   U^r + \sqrt{(U^r)^2 + f
		(r)} \sqrt{\beta^2 - \frac{f (r)}{r^2}} \right)}{\sqrt{1 + \left(
		\frac{r}{f} \left( - \beta   U^r + \sqrt{(U^r)^2 + f (r)}
		\sqrt{\beta^2 - \frac{f (r)}{r^2}} \right) \right)^2}} \nonumber\\
& = & \frac{- \frac{\beta   U^r}{\sqrt{f}} + \sqrt{1 +
		\frac{(U^r)^2}{f}} \sqrt{\beta^2 - \frac{f (r)}{r^2}}}{\beta \sqrt{1 +
		\frac{(U^r)^2}{f}} - \frac{U^r}{\sqrt{f}} \sqrt{\beta^2 - \frac{f}{r^2}}} \nonumber\\
& = & \frac{\sqrt{\beta^2 - \frac{f (r)}{r^2}} + \left( \beta +
	\sqrt{\beta^2 - \frac{f (r)}{r^2}} \right) \left( \frac{(U^r)^2}{f} -
	\frac{U^r}{\sqrt{f}} \sqrt{1 + \frac{(U^r)^2}{f}} \right)}{\left( \beta
	\sqrt{1 + \frac{(U^r)^2}{f}} - \frac{U^r}{\sqrt{f}} \sqrt{\beta^2 -
		\frac{f}{r^2}} \right) \left( \sqrt{1 + \frac{(U^r)^2}{f}} -
	\frac{U^r}{\sqrt{f}} \right)} \nonumber\\
& = & \cos \alpha_{\rm{radial}}~.
\end{eqnarray}
We start from the angular distance that we calculated in Lemaitre coordinate and then derive the results from Lebedev and Lake \cite{lebedev_influence_2013}. It shows that $\psi_{\rm mov} = \alpha_{\rm radial}$. 


\bibliography{citation}
\end{document}